\documentclass[preprint]{revtex4}
\input epsf
\usepackage{amsmath,amssymb}
\usepackage{graphicx}

\draft

\begin{document}
\titlepage
\title{Matter loops corrected modified gravity in Palatini formulation}
\author{Xin-He Meng$^{1,2,3}$ \footnote{xhmeng@phys.nankai.edu.cn}
 \ \ Peng Wang$^1$ \footnote{pewang@eyou.com}
} \affiliation{1.  Department of Physics, Nankai University,
Tianjin 300071, P.R.China \\2. Institute of Theoretical Physics,
CAS, Beijing 100080, P.R.China \\3. Department of Physics,
University of Arizona, Tucson, AZ 85721}

\begin{abstract}
Recently, corrections to the standard Einstein-Hilbert action are
proposed to explain the current cosmic acceleration in stead of
introducing dark energy. In the Palatini formulation of those
modified gravity models, there is an important observation due to
Arkani-Hamed: matter loops will give rise to a correction to the
modified gravity action proportional to the Ricci scalar of the
metric. In the presence of such term, we show that the current
forms of modified gravity models in Palatini formulation,
specifically, the $1/R$ gravity and $\ln R$ gravity, will have
phantoms. Then we study the possible instabilities due to the
presence of phantom fields. We show that the strong instability in
the metric formulation of $1/R$ gravity indicated by Dolgov and
Kawasaki will not appear and the decay timescales for the phantom
fields may be long enough for the theories to make sense as
effective field theory . On the other hand, if we change the sign
of the modification terms to eliminate the phantoms, some other
inconsistencies will arise for the various versions of the
modified gravity models. Finally, we comment on the universal
property of the Palatini formulation of the matter loops corrected
modified gravity models and its implications.

\end{abstract}

\maketitle

It now seems well-established that the expansion of our universe is
currently in an accelerating phase. The most direct evidence for this is
from the measurements of type Ia supernova \cite{Perlmutter}.
Other indirect evidences such as the observations of CMB by the
WMAP satellite \cite{Spergel}, large-scale galaxy surveys by 2dF
and SDSS \cite{hui} also seem supporting this.

But now the mechanisms responsible for this acceleration are not
very clear. Many authors introduce a mysterious cosmic fluid
called dark energy to explain this (see Refs.\cite{Peebles,
Carroll-de, Padmanabhan} for a review). On the other hand, some
authors suggest that maybe there does not exist such mysterious
dark energy, but the observed cosmic acceleration is a signal of
our first real lack of understanding of gravitational physics
\cite{Lue}. An example is the DGP model \cite{Dvali}. Recently,
there are active discussions in this direction by modifying the
action for gravity [8-22]. Specifically, a $1/R$ term is suggested
to be added to the action \cite{CDTT, Capozziello}: the so called
$1/R$ gravity. It is interesting that such term may be predicted
by string/M-theory \cite{Odintsov-string}. Then the modifications
by adding $1/R+R^2$ \cite{Odintsov-R2} or $\ln R$ terms
\cite{Odintsov-lnR} are also suggested. For $1/R$ gravity, Vollick
\cite{Vollick} used Palatini variational principle to derive the
field equations. The modified Friedmann equations in Palatini
formulation and their properties are discussed in our previous
papers \cite{Wang1}. The Palatini formulation of the $1/R+R^2$ and
$\ln R$ theory are discussed in Ref.\cite{Wang3} and
Ref.\cite{Wang2} respectively. Flanagan \cite{Flanagan} derived
the equivalent scalar-tensor theory of the Palatini formulation.
Furthermore, in Ref.\cite{Flanagan2}, Flanagan derived the
equivalent scalar-tensor theory of a more general modified gravity
framework. His results provide a fundamental and powerful tool for
discussing modified gravity models in Palatini formulation. In the
following discussions, we will use the sign conventions for
various quantities in Flanagan's papers.

In general, when handled in Palatini formulation, one considers
the action to be a functional of the metric $\bar{g}_{\mu\nu}$ and
a connection $\hat{\bigtriangledown}_{\mu}$ which is independent
of the metric. The resulting modified gravity action can be
written as
\begin{equation}
S[\bar{g}_{\mu\nu},
\hat{\bigtriangledown}_{\mu}]=\frac{1}{2\kappa^2}\int
d^4x\sqrt{-\bar{g}}L(\hat{R}),\label{1.1}
\end{equation}
where $\kappa^2=8\pi G$,
$\hat{R}=\bar{g}^{\mu\nu}\hat{R}_{\mu\nu}$ and $\hat{R}_{\mu\nu}$
is the Ricci tensor of the connection
$\hat{\bigtriangledown}_{\mu}$. All the current existed forms of
$L$ can be written as $L(\hat{R})=\hat{R}+f(\hat{R})$. For $1/R$
gravity, $f(\hat{R})=-\alpha^2/(3\hat{R})$ \cite{CDTT, Vollick};
for $1/R+R^2$ gravity,
$f(\hat{R})=-\alpha^2/(3\hat{R})+\hat{R}^2/(3\beta)$
\cite{Odintsov-R2, Wang3}; for $\ln R$ gravity,
$f(\hat{R})=\beta\ln(\hat{R}/\alpha)$ \cite{Odintsov-lnR, Wang2}.
As shown by Flanagan \cite{Flanagan}, these theories, when written
in a canonical form, contain a scalar field with no kinetic term.
Thus they can be seen as General Relativity coupled to modified
matter actions \cite{Flanagan2}.

The starting point of the current work is an observation due to
Nima Arkani-Hamed (see Ref.\cite{Flanagan2} for a more detailed
discussion of Arkani-Hamed's idea): the theory (\ref{1.1}) has
fine-tuning problems as an effective quantum field theory.
Specifically, matter loops will give rise to a correction to the
action (\ref{1.1}) proportional to the Ricci scalar $\bar{R}$ of
the metric $\bar{g}_{\mu\nu}$. Thus in this paper, we will study
the following matter loops corrected modified gravity action
\begin{equation}
S_{loop}[\bar{g}_{\mu\nu},
\hat{\bigtriangledown}_{\mu}]=\frac{1}{2\kappa^2}\int
d^4x\sqrt{-\bar{g}}[\bar{R}+f(\hat{R})]\equiv
S_{EH}+S_{Palatini},\label{1.2}
\end{equation}
where $S_{EH}$ is the familiar Einstein-Hilbert action. When
written in the canonical form, the presence of a $\bar R$ term
will induce a kinetic term for the scalar field. Note that for
simplicity of discussions, we ignored the linear $\hat{R}$ term.
Since the main motivation of modified gravity models is to explain
the current cosmic acceleration, the $f(\hat{R})$ term necessarily
dominates over the $\hat{R}$ term in realistic applications of the
models. Thus we believe it is sufficient and more illuminating to
consider only the combined effects of the induced Ricci scalar
$\bar{R}$ and $f(\hat{R})$. However, it can also be checked
explicitly that the inclusion of a linear $\hat R$ term will not
change the main conclusions drawn from action (\ref{1.2}).

Following Refs.\cite{Chiba, Odintsov-string}, we can write the
action $S_{Palatini}$ in an equivalent form. Introduce a scalar
field $\varphi$ and define the action
\begin{equation}
S_{\varphi}[\bar{g}_{\mu\nu}, \hat{\bigtriangledown}_{\mu},
\varphi]=\frac{1}{2\kappa^2}\int
d^4x\sqrt{-\bar{g}}[f'(\varphi)(\hat{R}-\varphi)+f(\varphi)].\label{2.1}
\end{equation}
Then if $f''(\hat{R})\neq 0$, the equation of motion of $\varphi$
implies $\varphi=\hat{R}$. Thus action (\ref{2.1}) is classically
equivalent to the action $S_{Palatini}$ (Unless indicated
explicitly, when we speak of two actions equivalent, we always
mean equivalence in classical level).

Define
\begin{equation}
\varepsilon=\text{sign} f'(\varphi),\label{}
\end{equation}
and define the scalar field $\sigma$ by
\begin{equation}
f'(\varphi)=\varepsilon e^{-2\sigma}.\label{2.3}
\end{equation}

Then a slight generalization of Flanagan's derivation
\cite{Flanagan} gives that the action (\ref{2.1}) is equivalent to
\begin{equation}
\tilde{S}[\tilde{g}_{\mu\nu}, \sigma]=\frac{1}{2\kappa^2}\int
d^4x\sqrt{-\tilde{g}}[\varepsilon\tilde{R}-V(\sigma)],\label{2.4}
\end{equation}
where
\begin{equation}
\tilde{g}_{\mu\nu}=e^{-2\sigma}\bar{g}_{\mu\nu},
\\ V(\sigma)=\varepsilon
e^{2\sigma}\varphi-e^{4\sigma}f(\varphi),\label{}
\end{equation}
and $\varphi$ is given in terms of $\sigma$ through
Eq.(\ref{2.3}). Thus, when considering the theory (\ref{1.1}),
with $f$ replaced by $L$ in the above expressions, in order to
have a positive coefficient of the Ricci scalar, we must have
$\text{sign}L'=1$. This is just the case for the Palatini
formulation of $1/R$ \cite{Vollick, Wang1} and $\ln R$ theory
\cite{Wang2}, where we have $\varepsilon=1$, so that
$\text{sign}L'=\text{sign}(1+f')=1$ is obviously satisfied.

Also transform the Einstein-Hilbert action $S_{EH}$ to the
$\tilde{g}_{\mu\nu}$ frame, and add it with action (\ref{2.4}),
the action (\ref{1.2}) is equivalent to
\begin{equation}
\tilde{S}_{loop}[\tilde{g}_{\mu\nu},
\sigma]=\frac{1}{2\kappa^2}\int
d^4x\sqrt{-\tilde{g}}[(e^{2\sigma}+\varepsilon)\tilde{R}+6e^{2\sigma}(\tilde{\bigtriangledown}\sigma)^2-
V(\sigma)].\label{2.6}
\end{equation}

Define
\begin{equation}
\eta=\text{sign}(e^{2\sigma}+\varepsilon),\label{}
\end{equation}
and define the scalar field $\chi$ as
\begin{equation}
e^{2\sigma}+\varepsilon=\eta e^{-2\chi}.\label{2.7}
\end{equation}

Let us conformally transform the metric as
\begin{equation}
\breve{g}_{\mu\nu}=e^{-2\chi}\tilde{g}_{\mu\nu}.\label{}
\end{equation}
Then the action (\ref{2.6}) is transformed to \cite{private}
\begin{equation}
\breve{S}_{loop}[\breve{g}_{\mu\nu}, \chi]=\frac{1}{2\kappa^2}\int
d^4x\sqrt{-\breve{g}}[\eta\breve{R}-6\eta(\breve{\bigtriangledown}\chi)^2+\frac{6e^{-2\chi}}{\eta
e^{-2\chi}-\varepsilon}
(\breve{\bigtriangledown}\chi)^2-\breve{V}(\chi)],\label{2.72}
\end{equation}
where $\breve{V}(\chi)=e^{4\chi}V(\sigma)$ and $\sigma$ are given
in terms of $\chi$ through Eq.(\ref{2.7}).

Now we can see from the action (\ref{2.72}) that, since the sign
of the coefficient of the Ricci scalar should be positive, we must
have $\eta=1$. Then, in order that the $\chi$ field have a right
sign for the kinetic term, we must have $\varepsilon=-1$. Thus
when adding the matter loops induced Ricci scalar into the action,
the $1/R$ gravity and $\ln R$ gravity in Palatini formulation
studied in Refs.\cite{Vollick, Wang1, Wang2}, where
$\varepsilon=1$, will have phantoms or negative energy
excitations, which may make the theory unstable classically
or(and) quantum mechanically. However, it is now well-known that
the presence of phantom fields does not necessarily mean that such
models is excluded as viable cosmological models. As shown
explicitly in Ref.\cite{Carroll-EOS} by field theoretic
computations, the decay timescale for the phantom field with a
gaussian potential can be long enough for the theory to make sense
as an effective field theory (see also Ref.\cite{Odintsov-phantom,
Piao} for more works on phantom cosmology). Below we will argue
that this is also hold for the matter loops corrected $1/R$
gravity in Palatini formulation. Before that, let us first
consider another sort of instability present in the metric
formulation of the $1/R$ gravity as indicated by Dolgov and
Kawasaki \cite{Dolgov}, which is much more serious than the
phantom decay instability. We will show that, for matter loops
corrected $1/R$ theory in Palatini formulation, this sort of
instability will not appear.

For the matter loops corrected $1/R$ theory,
$f(\hat{R})=-\alpha^2/(3\hat{R})$, the contracted field equations
are (one can derive those equations with a slight generalization
of the framework in Ref.\cite{Vollick})
\begin{equation}
-\bar{R}+\frac{\alpha^2}{\hat{R}}=\kappa T,\label{2.8}
\end{equation}
and
\begin{equation}
\hat{R}=\bar{R}+3(f')^{-1}\bar\bigtriangledown_0\bar\bigtriangledown_0f'-\frac{3}{2}(f')^{-2}
(\bar\bigtriangledown_0f')^2.\label{}
\end{equation}
Adding them together we can get
\begin{equation}
6\bar\bigtriangledown_0\bar\bigtriangledown_0\hat{R}-\frac{12}{\hat{R}}(\bar\bigtriangledown_0\hat{R})^2+\hat{R}^2
+\kappa T\hat{R}=\alpha^2.\label{}
\end{equation}
Let us consider the internal solution with time dependent matter
density as in Ref.\cite{Dolgov}. The first order correction to the
curvature $\hat{R}=\hat{R}_0+\hat{R}_1$ satisfies the equation
\begin{eqnarray}
6\bar\bigtriangledown_0\bar\bigtriangledown_0\hat{R}_1-\frac{24\bar\bigtriangledown_0\hat{R}_0}{\hat{R}_0}
\bar\bigtriangledown_0\hat{R}_1+(12(\frac{\bar\bigtriangledown_0\hat{R}_0}{\hat{R}_0})^2+2\hat{R}_0+\kappa
T)\hat{R}_1=\\\nonumber-6\bar\bigtriangledown_0\bar\bigtriangledown_0\hat{R}_0+\frac{12(\bar\bigtriangledown_0\hat{R}_0)^2}{\hat{R}_0}
-\hat{R}_0^2-\kappa T\hat{R}_0+\alpha^2. \label{2.11}
\end{eqnarray}

Since we expect $\hat{R}_0\sim -\kappa T$ and $T<0$, the
coefficient of $\hat{R}_1$ is positive in Eq.(16). Moreover, since
$\hat{R}$ and $\bar{R}$ are related to each other through the
algebraic equation (\ref{2.8}), thus the matter loops corrected
$1/R$ theory will not suffer from the instability of the sort
indicated by Dolgov and Kawasaki \cite{Dolgov}, which is due to a
large negative coefficient of the $\hat{R}_1$ term. The absence of
this sort of instability also happens in the Palatini formulation
of the $1/R$ theory without matter loops induced Ricci scalar
\cite{Wang1}.

Then we turn to the analysis of whether the decay timescales for
the phantom fields can be long enough for the theories to make
sense as effective field theory \cite{Carroll-EOS}. We still
consider the matter loops corrected $1/R$ theory as an
illuminating example. The potential $\breve{V}$ is given by
\begin{equation}
\breve{V}(\chi)=e^{4\chi}V(\sigma)=\frac{2\alpha}{\sqrt3}e^{4\chi}(e^{-2\chi}-1)^{\frac{3}{2}}.\label{}
\end{equation}

To apply the analysis of Ref.\cite{Carroll-EOS}, we need first to
transform the kinetic term in action (\ref{2.72}) to the standard
form of phantom field. Define $\Phi=M_{Pl}\text{arcsin}(e^\chi)$,
then action (\ref{2.72}) with $\eta=\epsilon=1$ can be rewritten
as:
\begin{equation}
\breve{S}_{loop}[\breve{g}_{\mu\nu}, \Phi]=\frac{1}{2\kappa^2}\int
d^4x\sqrt{-\breve{g}}\breve{R}-\int
d^4x\sqrt{-\breve{g}}[-3(\bigtriangledown\Phi)^2+\bar
V(\Phi)],\label{}
\end{equation}
where
\begin{equation}
\bar
V(\Phi)=\frac{M_{pl}^2\alpha}{\sqrt3}\sin(\frac{\Phi}{M_{Pl}})\cos^3(\frac{\Phi}{M_{Pl}}).\label{po}
\end{equation}
 Note since $\exp(-2\chi)=\exp(2\sigma)+1$, we have
$\chi<0$, then $0<\Phi<\pi/2$. See Fig.1.

\begin{figure}
  \includegraphics[width=0.8\columnwidth]{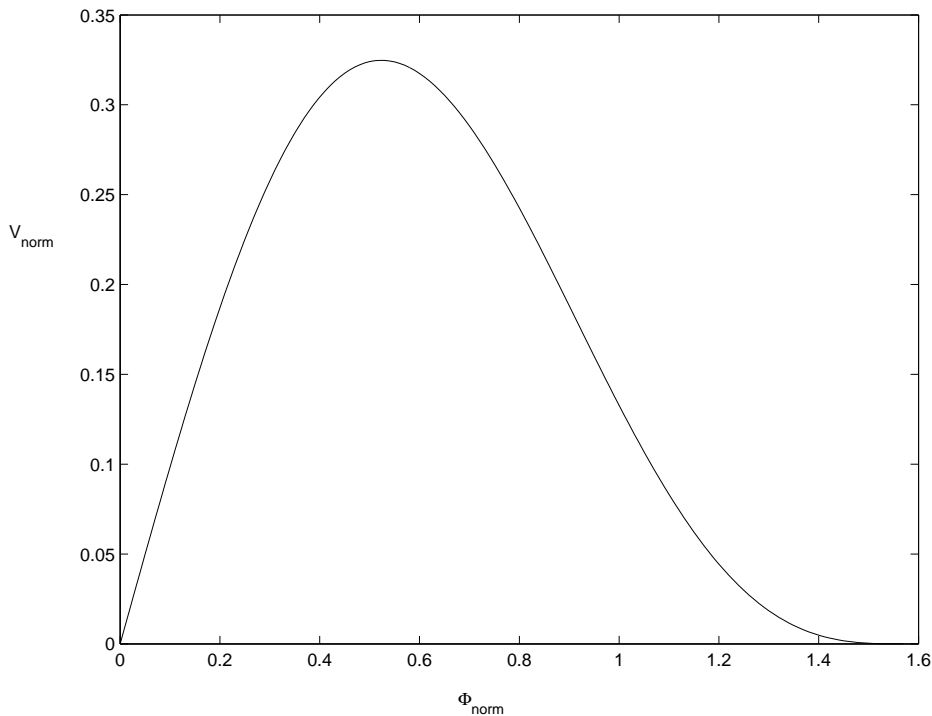}
  \caption{The scalar potential given by Eq.(\ref{po}). $\Phi_{norm}\equiv\Phi/M_{Pl}$ and $V_{norm}\equiv
  \sqrt3V/(M_{Pl}^2\alpha)$.}\label{1}
\end{figure}

Then we will follow closely the analysis of
Ref.\cite{Carroll-EOS}. But in the current case, there is a
difference with the case considered in Ref.\cite{Carroll-EOS}:
here the phantom field is conformally coupled to matter and thus
phantoms may decay to matter particles. We will ignore those decay
channels in the present analysis and restrict our attentions to
only gravitons and phantoms. Then, the phantom decay channel
involving the smallest number of particles is
\begin{equation}
\Phi\longrightarrow h+\Phi_1+\Phi_2\label{decay}
\end{equation}
where $h$ is a graviton and $\Phi,\Phi_1,\Phi_2$ is phantom
particles.

Since scalar fields with negative kinetic terms evolve to the
maxima of their classical potential, where from Eq.(\ref{po}) and
Fig.1 we can see $\Phi_{max}\sim M_{Pl}$ and $V(\Phi_{max})\sim
M_{Pl}^2\alpha$. Thus we will consider the potential (\ref{po})
expanded as a power series around the background value
$\Phi_{max}$. To study the decay (\ref{decay}), the interaction
part of the Lagrangian is required to be first order in $h$ and
third order in $\Phi$, which is
\begin{equation}
L_I=\frac{1}{M_{Pl}}(M_{Pl}h)\frac{1}{3!}V'''(\Phi_{max})\Phi^3=\lambda_{eff}(M_{Pl}h)\Phi^3.\label{}
\end{equation}
The effective coupling constant can be estimated as
\begin{equation}
\lambda_{eff}\sim\frac{V(\Phi_{max})}{M_{Pl}^4}\sim\frac{\alpha}{M_{Pl}^2}\sim
10^{-120},\label{}
\end{equation}
where we have used in the last the step the condition: $\alpha\sim
(10^{-33}eV)^2$, see Refs.\cite{CDTT, Vollick}. This is just the
same order of magnitude as in Ref.\cite{Carroll-EOS}. Then the
remaining part of the analysis is the same as in
Ref.\cite{Carroll-EOS}: such a small coupling constant will lead
to a decay timescale that can be larger than the Hubble time;
moreover, the total decay timescale remains of order of
(\ref{decay}) as long as the momentum cut-off is not larger than
$M_{Pl}$, which is a reasonable assumption. Thus we can conclude
that the decay timescale for the phantom field with potential
(\ref{po}) can also be long enough for the theory to make sense as
an effective field theory.

Next, we will explore whether one can modify the current modified
gravity models to eliminate the phantom field, i.e.
$\varepsilon=-1$, without other inconsistencies.

For the $1/R$ theory, if one changes the sign of the action to be
$f(\hat{R})=\alpha^2/(3\hat{R})$ so that $\varepsilon=-1$, another
more serious problem arises. The vacuum field equation for such an
action is
\begin{equation}
-\bar{R}-\frac{\alpha^2}{\hat{R}}=0,\label{2.12}
\end{equation}
and the equation by varying the connection will give
$\bar{R}=\hat{R}$. Thus there is no real solution of the vacuum
equation (\ref{2.12}). This obviously can not be a physical
theory.

For the $\ln R$ theory, if one changes the sign of the action to
be $f(\hat{R})=-\beta\ln(\hat{R}/\alpha)$ so that
$\varepsilon=-1$. The situation is not as serious as the $1/R$
theory. In the $\epsilon=1$ case, there is an unique
correspondence between $\hat{R}$ and any constant value of $T$;
while in the $\epsilon=-1$ case, the correspondence might be
two-to-one, or no solution for a given constant $T$, depending on
the value $\alpha/\beta$, but never one-to-one (see the discussion
in Ref.\cite{Odintsov-lnR} for more details on this case in metric
formulation). This is also not a good sign for the theory to
become a physically promising candidate theory.

For the $R^2$ theory, whose Palatini formulation is studied in
Ref.\cite{Wang3}, one should take $f(\hat{R})=-\hat{R}^2/\beta$,
where $\beta>0$. Now the situation is better, the correspondence
between $\bar{R}$ and any constant $T$ is still one-to-one. On the
other hand, there is a constraint on $\hat{R}$: from
Eq.(\ref{2.7}), $\sigma>0$; and from Eq.(\ref{2.3}),
$2\hat{R}/\beta=e^{-2\sigma}$. Thus this theory can be consistent
only when $\hat{R}<\beta/2$. The parameter $\beta$ can be
constrained by the BBN data, see Refs.\cite{Carroll-BBN, Wang3}.

For a general $f(\hat R)$, from the vacuum equation
\begin{equation}
-\hat R_0+f'(\hat R_0)\hat R_0-2f(\hat R_0)=0,\label{}
\end{equation}
where we have substituted the relation $\bar R=\hat R$ followed
from the equation of motion by varying the connection, one can see
that if one wants the theory to be free of phantom, i.e. $f'(\hat
R_0)<0$, and the vacuum equation has real positive solution, one
must require $f(\hat R_0)<0$. Thus all the realistic value of
$\hat R$ in the presence of matter should satisfy $f(\hat R)<0$
since $f'(\hat R)<0$. But in order to drive a recent acceleration,
$f\longrightarrow +\infty$ while $\hat R\longrightarrow0$. All the
current forms of $f(\hat R)$ do not satisfy those conditions all.
Actually, we strongly suspect that a reasonable $f(\hat R)$
satisfying all those conditions exists. Thus, one have to
accept the presence of phantom in the Palatini formulation of
matter loops corrected modified gravity models, which is not so
catastrophic as we have shown in this paper.

An important feature of the Palatini formulation of modified
gravity is that it will give the usual vacuum Einstein equations
for generic Lagrangian of the form $L(R)$: the so called
'universal property' of Palatini formulation of modified gravity
model \cite{Volovich}. It can be easily shown that this feature is
still hold for the action (\ref{1.2}). Thus in classical level,
theories of the type (\ref{1.1}) and (\ref{1.2}) are actually not
modifications of gravity at all. As can be checked explicitly, the
weak field expansions around a background spacetime (Minkowskii,
de Sitter and Anti-de Sitter) will give almost the same equations as for
General Relativity and the only difference is in the source term.
This means the behavior of graviton is
unchanged. Furthermore, as shown in \cite{Rubilar}, when the trace
of the energy-momentum tensor, $T$, is a constant value, the
Palatini formulation of theory (\ref{1.1}) still reduces to
General Relativity. It is easy to show that this is also hold for
the theory (\ref{1.2}). The key point is that for vacuum or
constant $T$, from the equation of motion by varying the
connection, we have $\hat R_{\mu\nu}=\bar R_{\mu\nu}$. This means
that the non-relativistic gravitational potential generated by a
static source is the standard Newtonian one. Also can be checked
explicitly, the potential for a slowly moving object behaves still
as $\sim 1/r$, while the nominator contains not only the mass, but
also the first and second order derivatives of the trace $T$.
Thus, roughly speaking, in the Palatini formulation, spacetime
curvature depends not only on the value of energy-momentum tensor,
but also its first and second order derivatives. The precise form
of the gravitational potential and a detailed discussion of the
above assertions will be presented elsewhere \cite{Wang4}. We also
comment that the modified gravity models will give several
predictions that are distinguishable with the dark energy models
\cite{Lue}, including the power spectrum of large scale structure, the ISW effects,
etc. The observations of those quantities can be used to discriminate between
modified gravity models and dark energy models.

In concluding, we have shown that in the presence of a matter
loops induced Ricci scalar term, the current forms of modified
gravity models in Palatini formulation, specifically, the $1/R$
gravity and $\ln R$ gravity, will have phantoms. However, the
phantom will not lead to the strong instability in the metric
formulation of modified gravity and the decay timescale for the
phantom field may be long enough for the theory to make sense as
an effective field theory . Thus, based on the current analysis,
the $1/R$ and $\ln R$ gravity with an induced Ricci scalar term
are still viable cosmological models despite the presence of
phantom fields. On the other hand, the versions without phantoms
are unacceptable because of serious inconsistencies. Note that the
above conclusions are based on the classical equivalence of
various conformally transformed actions. Of course, on quantum
level it is well-known (see explicit examples for quantum
dilatonic and higher derivative gravities \cite{Odintsov5,
Buchbinder, Grumiller}) that even classically equivalent theories
are not always equivalent on quantum level.

\textbf{Acknowledgements}

We would especially like to thank Professor Sergei Odintsov for a
careful reading of the manuscript and many very helpful comments,
which have improved this paper greatly. We would like to thank
Professors A. Borowiec, \'{E}. \'{E}. Flanagan and S. Nojiri for
helpful discussions. We would also like to thank Professors D.
Grumiller and T. Padmanabhan to draw our attentions to their
interesting works. This work is partly supported by China NSF,
Doctoral Foundation of National Education Ministry and ICSC-World
lab. scholarship.

\end{document}